\documentclass[preprints,communication,accept,pdftex,moreauthors]{Definitions/mdpi} 
\usepackage[main = english, spanish, german]{babel}
\pdfoutput=1

\firstpage{1} 
\pubvolume{1}
\issuenum{1}
\articlenumber{0}
\pubyear{2024}
\copyrightyear{2024}
\datereceived{1 December 2023} 
\daterevised{8 January 2024} 
\dateaccepted{ } 
\datepublished{ } 
\hreflink{https://doi.org/} 

\usepackage{dcolumn}

\Title{Towards Precision Muonic X-ray Measurements of Charge Radii of Light Nuclei}

\TitleCitation{Towards Precision Muonic X-ray Measurements of Charge Radii of Light Nuclei}

\Author{ {Ben} 
 Ohayon~$^{1,}$*\orcidA{},
Andreas Abeln~$^{2}$,
Silvia Bara~$^{3}$\orcidC{},
Thomas Elias Cocolios~$^{3}$\orcidD{},
Ofir Eizenberg~$^{1}$\orcidU{},
Andreas Fleischmann~$^{2}$\orcidE{},
Loredana Gastaldo~$^{2}$\orcidF{},
C\'esar Godinho~$^{4,5}$\orcidG{},
Michael Heines~$^{3}$\orcidH{},
Daniel Hengstler~$^{2}$,
Guillaume Hupin~$^{5}$\orcidJ{},
Paul Indelicato~$^{6}$\orcidK{},
Klaus Kirch~$^{7,8}$\orcidL{},
Andreas Knecht~$^{8}$\orcidM{},
Daniel Kreuzberger~$^{2}$,
Jorge Machado~$^{4}$\orcidN{},
Petr Navratil~$^{9}$\orcidO{},
Nancy Paul~$^{6,}$*\orcidP{},
Randolf Pohl~$^{10,11}$\orcidQ{},
Daniel Unger~$^{2}$,
Stergiani Marina Vogiatzi~$^{8}$\orcidR{},
Katharina von Schoeler~$^{7,8}$\orcidS{}
and Frederik Wauters~$^{10,12}$\orcidT{}
}

\AuthorNames{Ben Ohayon,
Andreas Abeln,
Silvia Bara,
Thomas Elias Cocolios,
Ofir Eizenberg,
Andreas Fleischmann,
Loredana Gastaldo,
C\'esar Godinho,
Michael Heines,
Daniel Hengstler,
Guillaume Hupin,
Paul Indelicato,
Klaus Kirch,
Andreas Knecht,
Daniel Kreuzberger,
Jorge Machado,
Petr Navratil,
Nancy Paul,
Randolf Pohl,
Daniel Unger,
Stergiani Marina Vogiatzi,
Katharina von Schoeler,
Frederik Wauters}

\AuthorCitation{Ohayon, B.; Abeln, A.; Bara, S.; Cocolios, T.E.; Eizenberg, O.; Fleischmann, A.; Gastaldo, L.; Godinho, C.; Heines, M.; Hengstler, D.; Hupin, G.; Indelicato, P.; Kirch, K.; Knecht, A.; Kreuzberger, D.; Machado, J.; Navratil, P.; Paul, N.; Pohl, R.; Unger, D.; Vogiatzi, S.M.; 
 {von Schoeler, K.;} Wauters, F.}

\address{%
$^{1}$ \quad Physics  {Department}, 
Technion-Israel Institute of Technology, 
Haifa 3200003, Israel;  {ofir.eiz@campus.technion.ac.il} 
\\
$^{2}$ \quad { {Kirchhoff} 
 Institut f\"ur Physik, Universit\"at Heidelberg}, 
Im Neuenheimer Feld 227, 69120 Heidelberg, Germany; andreas.abeln@kip.uni-heidelberg.de (A.A.); andreas.fleischmann@kip.uni-heidelberg.de (A.F.); loredana.gastaldo@kip.uni-heidelberg.de (L.G.); daniel.hengstler@kip.uni-heidelberg.de (D.H.); daniel.kreuzberger@kip.uni-heidelberg.de (D.K.); daniel.unger@kip.uni-heidelberg.de (D.U.)\\
$^{3}$ \quad KU Leuven,  {Instituut} 
 voor Kern- en Stralingsfysica, 3001 Leuven, Belgium; silvia.bara@kuleuven.be (S.B.); thomas.cocolios@kuleuven.be (T.E.C.); michael.heines@kuleuven.be (M.H.)\\
$^{4}$ \quad { {Departamento} 
 de F\'{\i}sica da Faculdade de Ci\^{e}ncias e Tecnologia, Universidade Nova de Lisboa, Monte~da~Caparica,}  2892-516 Caparica, Portugal; c.godinho@campus.fct.unl.pt (C.G.); jfd.machado@fct.unl.pt~(J.M.)\\
$^{5}$ \quad Le laboratoire de physique des deux infinis Irène Joliot-Curie  {(IJCLab),} 
 CNRS/IN2P3, {Universit\'e Paris-Saclay}, 91405 Orsay, France; guillaume.hupin@ijclab.in2p3.fr\\
$^{6}$ \quad Laboratoire Kastler Brossel, Sorbonne Universit\'{e}, CNRS, ENS-PSL Research University, Coll\`{e}ge de France, Case 74, 4, Place Jussieu, 75005 Paris, France; paul.indelicato@lkb.upmc.fr\\
$^{7}$ \quad Institute for Particle Physics and Astrophysics, ETH Z\"urich,  {8093} 
Z\"urich, Switzerland; klaus.kirch@psi.ch~(K.K.); kvonschoeler@phys.ethz.ch (K.v.S.)\\
$^{8}$ \quad Paul Scherrer Institute,  {5232}  {Villigen--PSI,} 
 Switzerland; a.knecht@psi.ch (A.K.); stella.vogiatzi@psi.ch (S.M.V.)\\
$^{9}$ \quad  {TRIUMF}, 4004 Wesbrook Mall, Vancouver, BC V6T 2A3, Canada; navratil@triumf.ca\\
$^{10}$ \,\, {PRISMA+ Cluster of Excellence, Johannes Gutenberg-{Universit\"at}} Mainz, 55128 Mainz, Germany; pohl@uni-mainz.de (R.P.); fwauters@uni-mainz.de (F.W.)\\
$^{11}$ \,\, {Institut f\"ur Physik, {QUANTUM}, Johannes Gutenberg-{Universit\"at}} Mainz, 55128 Mainz, Germany\\
$^{12}$ \,\, {Institut {f\"ur} Kernphysik, Johannes Gutenberg-{Universit\"at}} Mainz, 55128 Mainz, Germany}

\corres{Correspondence: bohayon@technion.ac.il (B.O.); npaul@lkb.upmc.fr (N.P.)}

\abstract{We, the QUARTET collaboration, propose an experiment to measure the nuclear charge radii of light elements with up to 20~times higher accuracy. These are essential both for understanding nuclear physics at low energies, and for experimental and theoretical applications in simple atomic systems. 
Such comparisons advance the understanding of bound-state quantum electrodynamics and are useful for searching for new physics beyond the Standard Model.
The energy levels of muonic atoms are highly susceptible to nuclear structure, especially to the mean square charge radius.
The radii of the lightest nuclei (with the atomic number, $Z=1,2$) have been determined with high accuracy using laser spectroscopy in muonic atoms, while those of medium mass and above were determined using X-ray spectroscopy with semiconductor detectors.
In this communication, we present a new experiment, aiming to obtain precision measurements of the radii of light nuclei $3 \leq Z \leq 10$ using 
single-photon energy measurements with cryogenic microcalorimeters; a quantum-sensing technology capable of high efficiency with outstanding resolution for low-energy X-rays.
}

\keyword{muonic atoms; charge radius; X-ray; metallic magnetic calorimeter (MMC); nuclear structure; bound-state 
quantum electrodynamics (QED), Simple atomic systems} 

\begin{document}
\section{Introduction}

Muonic atoms are highly suitable systems for studying the nucleus. 
Due to the heavy mass of muons ($m_{\mu} \sim 200~m_{e}$, with $m_e$ the electron mass), the Bohr radius of muonic atoms is approximately 200 times smaller than that of electronic atoms, and thus, for low angular momentum states, the muon wavefunction has a $200^3 \approx $ 
$10^6$ times larger overlap with that of the nucleus. 
The nuclear properties thus lead to measurable shifts in the atomic transition energies, making muonic atom spectroscopy an effective probe of phenomena such as finite nuclear size effects~\cite{1969-mu,1979-Friar,2003-Friar,2005-PSAS,2018-HFS, 2020-FAMU, 2022-Muonic}, relativistic 
QED (quantum electrodynamics) contributions~\cite{2005-PSAS,2007-Eides,2022-Muonic,2022-HFS-rec}, and possible short-range interactions carried by new mediators~\cite{2010-SvaliNP,2011-protonNP,2018-Axion,2017-NP,2018-X17,2014-Force,2022-Clara}. These systems are particularly well-suited to accurately determining the RMS 
(root mean squre) nuclear charge radius (the slope of the Sachs form factor at small momentum transfer, henceforth
called `radius' for brevity), which can be obtained using the spectroscopy of low-lying radiative transitions (mostly $2P-1S$)~\cite{1969-mu,1974-Review,1995-Fricke}.
Indeed, historically, the best measurements of absolute radii have been obtained using muonic atom spectroscopy, sometimes leading to surprising results such as the `proton radius puzzle' 
~\cite{2012-proton,2013-Pohl,2015-proton,2020-proton,2021-proton,2022-proton,2023-proton}.  

The radius is a fundamental property of the nucleus, and knowledge of it is not only 
significant in the development of a nuclear structure theory, but also for obtaining a reliable comparison between the experimental results and theoretical expectations at the accuracy frontier of Standard Model tests with atoms and nuclei~\cite{2003-Friar}. Accordingly, the radius of the proton and deuteron are considered fundamental constants, on a similar footing as their masses~\cite{Codata2018}, with those of heavier nuclei expected to be included in the next CODATA adjustment of fundamental constants.

In Table~\ref{tab:rad}, we collect the most precise values for the radii of light and stable nuclei. It is immediately seen that the elements in the most critical need of improved radii are those with a nuclear charge, $Z$, just above helium, a region that is currently beyond the reach of laser spectroscopy, where solid-state detectors are the most unsuitable. Advantageously, these nuclei are also those whose structure can now be calculated by the most advanced ab initio nuclear theory methods, as detailed in what follows. 

\begin{table}[H] 
\caption{Current status of the charge radii {, $r_c$,} of light stable nuclei. The method abbreviations are as follows: '$\mu$-Laser'---muonic atoms laser spectroscopy; 'el. scat.'---elastic electron scattering; 'OIS'---optical isotope shift; '$\pi^+$ scat.'---elastic positive pion scattering; '$\mu$-X'---muonic atom X-ray spectroscopy.
Numbers in the parentheses mark the uncertainty to the preceding digits, which is also denoted by $\sigma_r$. } 
\label{tab:rad}
\begin{tabularx}{\textwidth}{lCClC}
\toprule
 	& \boldmath{$r_c$}\textbf{/fm}	& \boldmath{$\sigma_r r_c^{-1}/10^{-3}$} & \textbf{Method} & \textbf{Reference}\\
\midrule
$^1$H 	 & $0.84060(39)$   & 0.5 & $\mu$-Laser	  &\cite{2013-Antognini,2023-Theory}\\
$^2$H 	 & $2.12775(17)$   & 0.1 & OIS$~+~r_c$($^1$H) &\cite{2018-IS,2010-IS}\\
$^3$He 	 & $1.97007(94)$   & 0.5 & $\mu$-Laser	&\cite{2023-Theory,2023-Helion}\\
$^4$He 	 & $1.6786(12)$~ ~ & 0.7 & $\mu$-Laser	&\cite{2021-Krauth,2023-Theory}\\
$^6$Li 	 & $2.589(39)$~ ~ ~& 15 ~& el. scat.		&\cite{1967-LiScat,1972-LiScat,1971-LiScat,2011-Li} \\
$^7$Li 	 & $2.444(42)$~ ~ ~& 17 ~& OIS$~+~r_c$($^{6}$Li)	&\cite{2011-Li,2011-LiIS} \\
$^9$Be 	 & $2.519(32)$~ ~ ~& 13 ~& el. scat.~\textsuperscript{1}&\cite{1972-Be}\\
$^{10}$B & $2.510(31)$~ ~ ~& 12 ~& OIS$~+~r_c$($^{11}$B) &\cite{2019-B} \\
$^{11}$B & $2.411(21)$~ ~ ~& 8.7 & $\pi^+$ scat.$~+~r_c$($^{12}$C) &\cite{1980-B}\\
$^{12}$C & $2.4829(19)$~ ~ & 0.8 & $\mu$-X &\cite{1984-Ruckstuhl}\\
$^{13}$C & $2.4628(39)$~ ~ & 1.6 & $\mu$-X &\cite{1985-13C}\\
\bottomrule
\end{tabularx}
\noindent{\footnotesize{\textsuperscript{1}
Estimated using a model-dependent analysis of an electron scattering experiment covering a narrow momentum transfer range. 
The same study quotes a radius for $^{12}$C~\cite{1972-Be}, which differs by three standard deviations from the  
modern value. A systematic uncertainty was added in quadrature to the smaller experimental uncertainty.
}}
\end{table}

\section{Physics Cases}
\label{sec:PhysicsCases}

\subsection{Nuclear Structure}
\label{sec:NucStruct}

In contemporary ab initio approaches, nucleon--nucleon and three-nucleon interactions are derived from chiral effective field theory (EFT) and used to calculate observables in quantum many-body systems with quantifiable uncertainties~\cite{2009-Epelbaum,2020-Hergert}. 
Due to a combinatorial increase in computational cost with mass number,
$A$, high-precision calculations are limited to  {$A \approx 16$.}
The next-generation calculations also treats the coupling to the continuum and what is referred to as ``open quantum systems''~\cite{2016-NCSMC, 2016-NAv, 2019-Vorrabi}, which are crucial for accurately reproducing the structure of both light and exotic nuclei, especially their spatial extension.

While nuclear forces are fit for the properties of light, dilute nuclear systems, testing their ability to predict the properties of $A>4$ 
bound systems (i.e., when nuclear density achieves saturation but $A$ remains sufficiently low for highly precise calculations) provides   
a means to gauge the measurements accuracy. The test allows for an investigation of the precision of the chiral EFT itself, ultimately challenging the understanding of the strong force at low energies. Nuclear radii are a particularly interesting testing ground, as they can be calculated to a high level of precision when the coupling with continuum is included, and the measurements of absolute charge radii may even be used to obtain, e.g., electric quadrupole (E2) observables by applying the observed correlations~\cite{2022-Caprio}.  

As an example, we consider the radii of the lithium isotopic chain, whose isotopic differences relative to $^6$Li were extracted with high precision from optical isotope shift measurements~\cite{2004-Li89, 2006-Li911,2011-Li}.
Different ab initio nuclear models have been put forward to reproduce the results. However, the results cannot distinguish them, because of the dominating uncertainty in the $^6$Li radius, to which the chain is referenced. 
The same happens in the beryllium \cite{2009-Be} and boron \cite{2019-B} {cases}. 
The corresponding measurement goal is to distinguish well which model reproduces the measured radii within a few  {$10^{-3}$,} of its value, which can be reached at the early stage of an experiment.

A more demanding accuracy is required when considering differential observables, such as the differences between the radii of mirror nuclei (the nuclei with neutron and proton numbers being interchanged): 
$\Delta r_\mathrm{mir}$, which are a focus of contemporary studies of  
nuclear structure (see e.g.,~\cite{2017-Brown, 2020-ArCa, 2021-Pineda, 2022-ISO, 2022-info, 2023-MirrStellar}).
A linear relationship between $\Delta r_\mathrm{mir}$ and neutron skins~\cite{2018-Mirr, 2020-Mirr, 2023-Mirr}, which are particularly complicated to directly measure, has been found. Measuring $\Delta r_\mathrm{mir}$ can thus contribute to the understanding of the variations in neutron skin with isospin, while any deviations from a linearity would
indicate the role that continuum degrees of freedom play in exotic nuclei structure evolution. Light mirror pairs such as $^7$Li-$^7$Be and $^8$Li-$^8$B possess a large isospin asymmetry and are, hence, well-suited to testing these theoretical predictions. 
Differences in radii were measured with the optical isotope shifts in the Li~\cite{2004-Li89, 2006-Li911,2011-Li} and Be~\cite{2009-Be} chains, while the measurement of  $^8$B~\cite{2017-B8} is ongoing. A need to measure $\Delta r_\mathrm{mir}$ accurately enough to not be limited by the reference radii sets a stringent demand to measure the radii 
with the accuracy as high as $10^{-3}$ for a stable isotope of each of the three elements.

\subsection{QED and Beyond Standard Mode} 
\label{sec:QEDaBSM}

The cases discussed in Section \ref{sec:NucStruct}, are rooted in nuclear physics. Here, we show that high-precision measurements of the radii are necessary for testing the QED effects at the frontier of research in atomic physics.
There are two main approaches to performing precision atomic structure calculations.
The first is based on a perturbative expansion with respect to relativistic and QED effects in the Coulomb field, with electron--electron correlations being treated non-perturbatively. This approach is best adapted for 
low-$Z$ systems such as hydrogen and helium.
The second approach treats the QED and relativistic effects of all orders in the Coulomb field. It is needed for high-$Z$ atoms, but loses accuracy in the low-$Z$ region, where the correlations are more pronounced. The intermediate region, $Z \approx 6$ is particularly interesting, as both approaches have each maximum uncertainty. There is therefore an interest in studying QED for that light, but not too light, few-electron systems (see, e.g.,~\mbox{\cite{2010-Yerokhin,2022-Yerokhin,2023-HLI}} and the discussion therein). 

Precise nuclear radii are needed to disentangle potentially missing QED contributions as a function of  {$Z$}. 
To illustrate this point, let us consider $2^3S_1-2^3P_j$ transitions in $^7$Li$^+$.
Their hyperfine-averaged value was determined to within $0.4\,$MHz~\cite{1994-Li}. From this measurement, the two-electron Lamb shift could be experimentally determined with $1.5\,$MHz precision, dominated by the uncertainty in the nuclear radius. Considering the ongoing experiments aim to obtain an accuracy of the order of $100\,$kHz~\cite{2020-LiOngoing}, an order-of-magnitude improvement in the radius would allow for the missing 
$\alpha^8$ theory contribution of the order $3\,$MHz, to be determined with an uncertainty of $10\%$~\cite{2023-HLI}.
Moreover, this effect can be scaled down to the analogous transition in the helium atom, and to shed light on the observed, and recently confirmed, deviations between experiment and theory~\cite{2021-He, 2023-He}.

Above Lithium, a new experimental program at Technical University Darmstadt, Germany, focuses on transitions in helium-like ions (HLIs) from beryllium to nitrogen~\cite{2020-BHLI, 2020-COALA,2023-HLIB, 2023-HeC}, with the measurements in $^{12}$C already completed~\cite{2023-HLIPRA, 2023-HLIPRL}.
Improved radii are crucial for confronting the results of the Darmstadt campaign with the state-of-the-art calculations, especially if one considers using measurements at a high $Z$ to determine the missing QED contributions, and using these calculations in the measurements with lower $Z$. 

At the accuracy frontier of beyond Standard Model (BSM) physics searches, new interactions are hunted through their manifestation as significant differences between experiment and theory.
Currently, the strongest bound on \textit{fifth forces} between charged leptons and neutrons derives from a combination of muonic and electronic isotope shifts in the hydrogen--deuterium pair~\cite{2017-NP}. When the mediating bosons are heavy, the sensitivity scale is $Z^3$, favoring highly charged systems~\cite{2022-Sailer}.
To utilize high-precision measurements of optical isotope shifts in simple enough electronic systems for BSM tests, one needs to considerably improve muonic isotope shifts.
Accordingly, and as both the nuclear theory and experimental uncertainties associated with calibration largely cancel in the difference, we consider to measure these isotope shifts with suitable precision to determine differential radii with an accuracy above $10^{-3}~$fm, limited by the residual nuclear theory uncertainty.

We conclude this Section by noting that improved measurements of transitions to the ground level in a muonic atom directly translate to a better prediction of the muonic atom Lamb shift, which is accessible to laser spectroscopy. This statement is true irrespective of the nuclear-structure uncertainty.
Quantitatively, a few parts per million measurement of the $2P-1S$ energy in muonic lithium translates to a few-meV prediction of the muonic lithium Lamb shift. Such a narrow search region greatly reduces the time needed to conduct a successful high-precision laser spectroscopy measurement, thus increasing the feasibility of the experiments suggested in Ref.~\cite{2018-LaserNext}.
The resolution afforded by laser spectroscopy in muonic atoms would, in turn, enable the hyperfine structure to be resolved and determine the Zemach radius (a convolution of the electric and magnetic distributions) of lithium isotopes. This determination is highly demanded  due to the sharp disagreement between this value, as calculated by nuclear theory and as determined by the electronic measurements~\cite{2008-Li,2023-LiHFS,2023-LiHFSTh,2023-LiMoments}.
Moreover, ongoing work suggests that the redundancy between X-ray and laser measurements in the same muonic species constitutes a powerful platform to search for new physics carried by new medium-mass (of the order of MeV) bosons.

\section{Theory Considerations}

The energy of the atomic transition between principle quantum numbers can be written as
\begin{equation}
    E=E_{\rm D} + \delta E_{\rm QED}+\delta E_{\rm FNS}+\delta E_{\rm TPE}+\cdots \, ,
\end{equation}
where $E_{\rm D}$ is the Dirac energy for a point-like nucleus, $\delta E_{\rm QED}$ is the sum of leading quantum electrodynamics effects, 
$\delta E_{\rm FNS}\propto r_c^2$ is the leading order correction due to the finite nuclear size, from which the charge radius is extracted, and $\delta 
E_{\rm TPE}$ is the sum of corrections stemming from the two-photon exchange, which depends on the nuclear structure, namely the nuclear polarizability and 
higher charge moments of the nucleus. 
At the precision level that is foreseen for the
project considered here, the uncertainty in point-nucleus QED corrections is negligible
~\cite{2016-LS_Light,2021-1S-2S}.
Accordingly, once the experimental accuracy of the transition energies is improved, the uncertainty in $\delta 
E_{\rm TPE}$ and higher-order nuclear structure contributions~\cite{2018-IS} is expected to dominate the derived radii. Based on the calculations for the lightest nuclei (see~\cite{2023-Theory} and references therein), preliminary results for $^{6,7}$Li~\cite{Muli2020} and the recent studies on heavier systems~\cite{2022-NP}, we estimate that a $5$ to $10\%$ uncertainty in the calculated $\delta E_{\rm TPE}$ is achievable, resulting in an absolute radii with an accuracy of the order of a few times $10^{-4}$, similar to that for the neighboring nuclei (see Table~\ref{tab:rad}).
Further calculations for the nuclei of interest are in progress.
These could be achieved by applying the no-core shell model~\cite{2013-NCSM} with the Lanczos method~\cite{1950-Lanc, 1974-Hay}.
In addition to the calculations, {the helpful} information on the nuclear shape can, in some cases (notably $^{12}$C~\cite{1982-ScatC}), be incorporated from elastic electron scattering measurements.

More accurate atomic theory considerations are needed in order to account for the unresolved fine and hyperfine structure features, mixed finite size and QED corrections, and shifts from spectator electrons, which screen the nuclear potential seen by the muon~\cite{1980-screen,1994-screen}.
Accordingly, we calculated the atomic structure of the targeted systems using the Multiconfiguration Dirac-Fock General Matrix Elements (MCDFGME) code, which is able evaluate the energies, transition probabilities, and hyperfine structure for exotic atoms composed of a nucleus, an arbitrary number of electrons, and an additional fermion or boson \cite{1975-Des, 1978-MDF, 2005-SPBI,2007-TAI,2013-Indel}.
The energies are obtained using a full-atomic wave function composed of a determinant with all the electrons, multiplied by the muon wave function, and by solving the full coupled system of differential equations. The electron--electron and muon--electron interactions are chosen to represent the full Breit operator with Coulomb, magnetic, and retardation in the Coulomb gauge. 
Nuclear deformation effects could also be of importance~\cite{1966-def,1967-def,1982-ZnDeform,2023-Zn,2023-def}. They are not explicitly included in the atomic structure calculation and will be evaluated separately for this work.

Based on successful studies in the lightest systems, we are confident that precision experiments with these heavier muonic atoms will instigate new activity in the relevant atomic and nuclear theories, potentially contributing to related fields, such as studies with antiprotonic atoms~\cite{2021-Exotic} and highly charged ions~\cite{2019-QEDHCI}.

\section{Experimental Considerations}
\label{sec:Experiment}

The radii of most of the stable nuclei were measured using traditional muonic X-ray spectroscopy with semiconductor detectors~\cite{2004-Fricke}. However, 
due to their moderate resolving power 
(fractional resolution) below 200~keV, compounded with the $Z^2$ scaling of the fractional contribution of the radius to the energy levels, this approach is insufficient to precisely determine finite-size effects in light nuclei. 
In contrast with the semiconducting detectors, crystal spectrometers offer high resolution in the multi-keV regime~\cite{1999-XRAY}. 
This detection method was used to determine the \mbox{$2P-1S$} transition energy in $\mu^{12}$C to 5~ppm, and derive the radius with an accuracy of $2\times10^{-3}\,$fm~\cite{1982-Ruckstuhl,1984-Ruckstuhl}. 
This demonstrates that an X-ray detector with a resolving power of a few thousand enables the radii measurements in light nuclei with a precision better than $10^{-3}$.
However, this method suffers from low efficiency and a narrow bandwidth, making it impractical to extend to the entire series of light muonic elements in the available facilities considering beamtime constraints.

In order to measure the relevant transitions in light elements with sufficient accuracy and within a reasonable time, we use a different technology: metallic magnetic calorimeters (MMCs)~\cite{2005-Fleischmann}, operated at 
exceptionally low cryogenic temperatures ($\sim 20\,$mK).
This quantum-sensing single-photon energy-detection technique achieves a high resolving power of few thousand, well-understood nonlinearity, and high quantum efficiency, as follows:
above 50\% up to $40~$keV, and above 5\% up to $180\,$keV~\cite{other}. It is thus ideally suited to tackling the problem of measuring the charge radii of light elements ($Z\leq10$) 
using X-ray spectroscopy in muonic atoms.
The principle of detection is that an X-ray is absorbed in a metallic absorber and its complete energy is converted to a temperature increase. This leads to a magnetization change in a paramagnetic material connecting the absorber to a thermal bath. This change in magnetization is high-sensitively detected using a superconducting quantum interference device (SQUID).

The first proof-of-concept measurement is currently undergoing its preparation. We transported an existing micro-calorimeter array for X-ray spectroscopy (maXs)~\cite{2015-MAXS, 2021-Unger} in a sidearm of a dilution refrigerator, from the Kirchhoff Institute for Physics in Heidelberg, to a secondary muon beamline at the Paul Scherrer Institute (PSI), Villigen, Switzerland.
The detector is planned to be integrated with the existing muon, electron, and photon detectors from the muX experiment~\cite{2021-MuX}, which is expected to allow identifying and suppressing different sources of background. 
For detailed information on the detector, its performance, and its integration with the beamline, see~\cite{other}. 

Precise absolute X-ray measurements not only require a high resolution, ample statistics, and effective background suppression, but also a robust calibration strategy.
The calibration function of the MMC detectors is to be determined periodically by using readily available sources~\cite{2020-MMCLin}.
%
To complement this method, a commercial X-ray tube is designed to be placed in the experimental system. There, electrons excite various metallic targets, emitting their characteristic X-rays, some of which have energies that are 
known to sub-ppm accuracy~\cite{2003-XRAY,2019-Mo}, and traceable to the SI-second~\cite{2020-SI}.
%
Similarly, the energy of muonic X-ray lines that do not involve an $S$-state can be reliably calculated to sub-ppm accuracy~\cite{2016-pion}.
In this way, higher-lying lines in heavier muonic atoms could calibrate $2P-1S$ lines in lighter systems.

Finally, let us note that the experiment considered in this paper and other next-generation experiments on exotic atoms (e.g.,~\cite{2023-Ne,2023-Daphne}), could significantly benefit from the ppm-level absolute gamma-ray energy 
measurements in the 20--200 keV range, especially from readily available long-lived commercial~sources.

\section{Summary}
QUARTET is a new Collaboration that seeks to significantly improve the experimental values of the charge radii from lithium to neon by means of a precision muonic X-ray spectroscopy with metallic magnetic calorimeters, filling the gap between laser spectroscopy- and semiconductor-based X-ray spectroscopy for {the} elements that are beyond reach of both the methods.
It will be the first time that those detectors are used with exotic atoms.
The expected ten-fold improvement in precision will significantly impact nuclear structure and QED tests, and pave the way for BSM physics searches through the combinations of the muonic (X-ray) and electronic (optical) isotope shifts.

\vspace{6pt} 


\funding{
B.O. is thankful for the support of the Council for Higher Education Program for Hiring Outstanding Faculty Members in Quantum Science and Technology.
%
The Kirchhoff Institute for Physics group at Heidelberg University is supported by Field Of Focus II initiative at Heidelberg University. D.U. acknowledges the support by the Research Training Group HighRR (GRK 2058) funded through the Deutsche Forschungsgemeinschaft, DFG.
The work of the KU Leuven group is supported by FWO-Vlaanderen (Belgium),  KU Leuven BOF C14/22/104, and European Research Council, grant no.~101088504 (NSHAPE).
%
P.N. acknowledges support from the NSERC Grant No. SAPIN-2022-00019. TRIUMF receives federal funding via a contribution agreement with the National Research Council of Canada.
%
The Lisboa group is supported in part by  {Funda\c{c}\~{a}o} 
ra a  {Ci\`encia} e Tecnologia (FCT; Portugal) through research center Grant No. UID/FIS/04559/2020 to LIBPhys-UNL.
The work of the ETH group was supported by the ETH Research Grant 22-2 ETH-023 {, Switzerland.}
}




\begin{adjustwidth}{-\extralength}{0cm}

\reftitle{References}

\PublishersNote{}
\end{adjustwidth}

\begin{thebibliography}{999}

\bibitem[Wu and Wilets(1969)]{1969-mu}
Wu, C.S.; Wilets, L.
\newblock Muonic atoms and nuclear structure.~{\em Annu. Rev. Nucl. Sci.} {\bf 1969}, {\em
  19},~527--606.
\newblock {\url{https://doi.org/10.1146/annurev.ns.19.120169.002523}.

\bibitem[Friar(1979)]{1979-Friar}
Friar, J.L.
\newblock Nuclear finite-size effects in light muonic atoms.
\newblock {\em Ann. Phys.} {\bf 1979}, {\em 122},~151--196.
 {https://doi.org/10.1016/0003-4916(79)90300-2} 

\bibitem[Friar(2003)]{2003-Friar}
Friar, J.L. The structure of light nuclei and its effect on precise atomic
  measurements.
\newblock In {\em Precision Physics of Simple Atomic Systems}; Karshenboim,
  S.G., Smirnov, V.B., Eds.; Springer: Berlin/Heidelberg, Germany, 2003; pp. 59--79.
\newblock {\url{https://doi.org/10.1007/978-3-540-45059-7_4}.

\bibitem[Karshenboim(2005)]{2005-PSAS}
Karshenboim, S.G.
\newblock Precision physics of simple atoms: QED tests, nuclear structure and
  fundamental constants.
\newblock {\em Phys. Rep.} {\bf 2005}, {\em 422},~1--63.
\newblock
  {\url{https://doi.org/10.1016/j.physrep.2005.08.008}}.

\bibitem[Kanda et~al.(2018)Kanda, Ishida, Iwasaki, Ma, Okada, Takamine, Ueno,
  Midorikawa, Saito, Wada, Yumoto, Oishi, Sato, Aikawa, Tanaka, and
  Matsuda]{2018-HFS}
Kanda, S.; Ishida, K.; Iwasaki, M.; Ma, Y.; Okada, S.; Takamine, A.; Ueno, H.;
  Midorikawa, K.; Saito, N.; Wada, S.;~et~al.
\newblock Measurement of the proton Zemach radius from the hyperfine splitting
  in muonic hydrogen atom.
\newblock {\em J. Phys. Conf. Ser.} {\bf 2018}, {\em
  1138},~012009.
\newblock {\url{https://doi.org/10.1088/1742-6596/1138/1/012009}}.

\bibitem[Pizzolotto et~al.(2020)Pizzolotto, Adamczak, Bakalov, Baldazzi,
  Baruzzo, Benocci, Bertoni, Bonesini, Bonvicini, Cabrera, et~al.]{2020-FAMU}
Pizzolotto, C.; Adamczak, A.; Bakalov, D.; Baldazzi, G.; Baruzzo, M.; Benocci,
  R.; Bertoni, R.; Bonesini, M.; Bonvicini, V.; Cabrera, H.;~et~al.
\newblock The FAMU experiment: Muonic hydrogen high precision spectroscopy
  studies.
\newblock {\em  Eur. Phys. J. A} {\bf 2020}, {\em 56},~185.
  {https://doi.org/10.1140/epja/s10050-020-00195-9}

\bibitem[Antognini et~al.(2022)Antognini, Bacca, Fleischmann, Gastaldo,
  Hagelstein, Indelicato, Knecht, Lensky, Ohayon, Pascalutsa, Paul, Pohl, and
  Wauters]{2022-Muonic}
Antognini, A.; Bacca, S.; Fleischmann, A.; Gastaldo, L.; Hagelstein, F.;
  Indelicato, P.; Knecht, A.; Lensky, V.; Ohayon, B.; Pascalutsa, V.;~et~al.
\newblock Muonic-atom spectroscopy and impact on nuclear structure and
  precision QED theory. \emph{ {arXiv} 
}{\bf 2022},  {arXiv:2210.16929.}
\newblock {\url{https://doi.org//10.48550/arXiv.2210.16929}}.

\bibitem[Eides et~al.(2007)Eides, Grotch, and Shelyuto]{2007-Eides}
Eides, M.I.; Grotch, H.; Shelyuto, V.A.
\newblock {\em Theory of Light Hydrogenic Bound States}; Springer
  Science \& Business Media:   {Berlin/Heidelberg, Germany,} 
   {2007.} 
  {https://doi.org/10.1007/3-540-45270-2} 

\bibitem[Antognini et~al.(2022)Antognini, Lin, and Meißner]{2022-HFS-rec}
Antognini, A.; Lin, Y.H.; Meißner, U.G.
\newblock Precision calculation of the recoil–finite-size correction for the
  hyperfine splitting in muonic and electronic hydrogen.
\newblock {\em Phys. Lett. B} {\bf 2022}, {\em 835},~137575.
\newblock
  {\url{https://doi.org/10.1016/j.physletb.2022.137575}}.

\bibitem[Karshenboim(2010)]{2010-SvaliNP}
Karshenboim, S.G.
\newblock Precision physics of simple atoms and constraints on a light boson
  with ultraweak coupling.
\newblock {\em Phys. Rev. Lett.} {\bf 2010}, {\em 104},~220406.
\newblock {\url{https://doi.org/10.1103/PhysRevLett.104.220406}}.

\bibitem[Barger et~al.(2011)Barger, Chiang, Keung, and Marfatia]{2011-protonNP}
Barger, V.; Chiang, C.W.; Keung, W.Y.; Marfatia, D.
\newblock Proton size anomaly.
\newblock {\em Phys. Rev. Lett.} {\bf 2011}, {\em 106},~153001.
\newblock {\url{https://doi.org/10.1103/PhysRevLett.106.153001}}.

\bibitem[Villalba-Ch\'avez et~al.(2018)Villalba-Ch\'avez, Golub, and
  M\"uller]{2018-Axion}
Villalba-Ch\'avez, S.; Golub, A.; M\"uller, C.
\newblock Axion-modified photon propagator, Coulomb potential, and Lamb shift.
\newblock {\em Phys. Rev. D} {\bf 2018}, {\em 98},~115008.
\newblock {\url{https://doi.org/10.1103/PhysRevD.98.115008}}.

\bibitem[Delaunay et~al.(2017)Delaunay, Frugiuele, Fuchs, and Soreq]{2017-NP}
Delaunay, C.; Frugiuele, C.; Fuchs, E.; Soreq, Y.
\newblock Probing new spin-independent interactions through precision
  spectroscopy in atoms with few electrons.
\newblock {\em Phys. Rev. D} {\bf 2017}, {\em 96},~115002.
\newblock {\url{https://doi.org/10.1103/PhysRevD.96.115002}}.

\bibitem[Jentschura and N\'andori(2018)]{2018-X17}
Jentschura, U.D.; N\'andori, I.
\newblock Atomic physics constraints on the $X$ boson.
\newblock {\em Phys. Rev. A} {\bf 2018}, {\em 97},~042502.
\newblock {\url{https://doi.org/10.1103/PhysRevA.97.042502}}.

\bibitem[Karshenboim et~al.(2014)Karshenboim, McKeen, and Pospelov]{2014-Force}
Karshenboim, S.G.; McKeen, D.; Pospelov, M.
\newblock Constraints on muon-specific dark forces.
\newblock {\em Phys. Rev. D} {\bf 2014}, {\em 90},~073004.
\newblock {\url{https://doi.org/10.1103/PhysRevD.90.073004}}.

\bibitem[Claudia and Clara(2022)]{2022-Clara}
Claudia, F.; Clara, P.
\newblock Muonic vs. electronic dark forces: A complete EFT treatment for atomic
  spectroscopy.
\newblock {\em J. High Energy Phys.} {\bf 2022}, {\em 2022},  {002.}
\newblock {\url{https://doi.org/10.1007/JHEP05(2022)002}}.

\bibitem[Engfer et~al.(1974)Engfer, Schneuwly, Vuilleumier, Walter, and
  Zehnder]{1974-Review}
Engfer, R.; Schneuwly, H.; Vuilleumier, J.; Walter, H.; Zehnder, A.
\newblock Charge-distribution parameters, isotope shifts, isomer shifts, and
  magnetic hyperfine constants from muonic atoms.
\newblock {\em At. Data Nucl. Data Tables} {\bf 1974}, {\em
  14},~509-- {597.} 
 {\url{https://doi.org/10.1016/S0092-640X(74)80003-3}}.

\bibitem[Fricke et~al.(1995)Fricke, Bernhardt, Heilig, Schaller, Schellenberg,
  Shera, and Dejager]{1995-Fricke}
Fricke, G.; Bernhardt, C.; Heilig, K.; Schaller, L.; Schellenberg, L.; Shera,
  E.; Dejager, C.
\newblock Nuclear ground state charge radii from electromagnetic interactions.
\newblock {\em At. Data Nucl. Data Tables} {\bf 1995}, {\em
  60},~177--285.
\newblock {\url{https://doi.org/10.1006/adnd.1995.1007}}.

\bibitem[Lorenz et~al.(2012)Lorenz, Hammer, and Mei{\ss}ner]{2012-proton}
Lorenz, I.; Hammer, H.W.; Mei{\ss}ner, U.G.
\newblock The size of the proton: Closing in on the radius puzzle.
\newblock {\em  Eur. Phys. J. A} {\bf 2012}, {\em 48},~151.

\bibitem[Pohl et~al.(2013)Pohl, Gilman, Miller, and Pachucki]{2013-Pohl}
Pohl, R.; Gilman, R.; Miller, G.A.; Pachucki, K.
\newblock Muonic hydrogen and the proton radius puzzle.
\newblock {\em Annu. Rev. Nucl. Part. Sci.} {\bf 2013}, {\em
  63},~175--204.
\newblock {\url{https://doi.org/10.1146/annurev-nucl-102212-170627}}.

\bibitem[Carlson(2015)]{2015-proton}
Carlson, C.E.
\newblock The proton radius puzzle.
\newblock {\em Prog. Part. Nucl. Phys.} {\bf 2015}, {\em
  82},~59--77.
\newblock {\url{https://doi.org/10.1016/j.ppnp.2015.01.002}}.

\bibitem[Hammer and Meißner(2020)]{2020-proton}
Hammer, H.W.; Meißner, U.G.
\newblock The proton radius: From a puzzle to precision.
\newblock {\em Sci. Bull.} {\bf 2020}, {\em 65},~257--258.
\newblock {\url{https://doi.org/10.1016/j.scib.2019.12.012}}.

\bibitem[Peset et~al.(2021)Peset, Pineda, and Tomalak]{2021-proton}
Peset, C.; Pineda, A.; Tomalak, O.
\newblock The proton radius (puzzle?) and its relatives.
\newblock {\em Prog. Part. Nucl. Phys.} {\bf 2021}, {\em
  121},~103901.
\newblock {\url{https://doi.org/10.1016/j.ppnp.2021.103901}}.

\bibitem[Jentschura(2022)]{2022-proton}
Jentschura, U.D.
\newblock Proton radius: A puzzle or a solution!?
\newblock {\em J. Phys. Conf. Ser.} {\bf 2022}, {\em
  2391},~012017.
\newblock {\url{https://doi.org/10.1088/1742-6596/2391/1/012017}}.

\bibitem[Meißner(2023)]{2023-proton}
Mei{\ss}ner, U.G.
\newblock The proton radius and its relatives---much ado about nothing?
\newblock {\em J. Phys. Conf. Ser.} {\bf 2023}, {\em
  2586},~012006.
\newblock {\url{https://doi.org/10.1088/1742-6596/2586/1/012006}}.

\bibitem[Tiesinga et~al.(2021)Tiesinga, Mohr, Newell, and Taylor]{Codata2018}
Tiesinga, E.; Mohr, P.J.; Newell, D.B.; Taylor, B.N.
\newblock CODATA recommended values of the fundamental physical constants:
  2018.
\newblock {\em Rev. Mod. Phys.} {\bf 2021}, {\em 93},~025010.
\newblock {\url{https://doi.org/10.1103/RevModPhys.93.025010}}.

\bibitem[Antognini et~al.(2013)Antognini, Nez, Schuhmann, Amaro, Biraben,
  Cardoso, Covita, Dax, Dhawan, Diepold, Fernandes, Giesen, Gouvea, Graf,
  Hänsch, Indelicato, Julien, Kao, Knowles, Kottmann, Le~Bigot, Liu, Lopes,
  Ludhova, Monteiro, Mulhauser, Nebel, Rabinowitz, dos Santos, Schaller,
  Schwob, Taqqu, Veloso, Vogelsang, and Pohl]{2013-Antognini}
Antognini, A.; Nez, F.; Schuhmann, K.; Amaro, F.D.; Biraben, F.; Cardoso,
  J.M.R.; Covita, D.S.; Dax, A.; Dhawan, S.; Diepold, M.;~et~al.
\newblock Proton structure from the measurement of 2S-2P transition frequencies
  of muonic hydrogen.
\newblock {\em Science} {\bf 2013}, {\em 339},~417--420.
\newblock {\url{https://doi.org/10.1126/science.1230016}}.

\bibitem[Pachucki et~al.(2022)Pachucki, Lensky, Hagelstein, Li~Muli, Bacca, and
  Pohl]{2023-Theory}
Pachucki, K.; Lensky, V.; Hagelstein, F.; Li~Muli, S.S.; Bacca, S.; Pohl, R.
\newblock Comprehensive theory of the Lamb shift in 
 { light muonic atoms.} 
\emph{ {arXiv}} {\bf 2022},  {arXiv:2212.13782}.
\newblock {\url{https://doi.org/10.48550/arXiv.2212.13782}}.

\bibitem[Pachucki et~al.(2018)Pachucki, Patk\'os, and Yerokhin]{2018-IS}
 {Pachucki, K.;} 
 Patk\'os, V.; Yerokhin, V.A.
\newblock Three-photon-exchange nuclear structure correction in hydrogenic
  systems.
\newblock {\em Phys. Rev. A} {\bf 2018}, {\em 97},~062511.
\newblock {\url{https://doi.org/10.1103/PhysRevA.97.062511}}.

\bibitem[Parthey et~al.(2010)Parthey, Matveev, Alnis, Pohl, Udem, Jentschura,
  Kolachevsky, and H\"ansch]{2010-IS}
Parthey, C.G.; Matveev, A.; Alnis, J.; Pohl, R.; Udem, T.; Jentschura, U.D.; Kolachevsky, N.; H\"ansch, T.W.~Precision measurement of the 
hydrogen-deuterium $1S-2S$ isotope shift.~{\em Phys. Rev. Lett.} {\bf 2010}, {\em 
104},~233001.~{\url{https://doi.org/10.1103/PhysRevLett.104.233001}}.

\bibitem[Schuhmann et~al.(2023)Schuhmann, Fernandes, Nez, Ahmed, Amaro, Amaro,
  Biraben, Chen, Covita, Dax, et~al.]{2023-Helion}
 {Schuhmann, K.; 
~et~al. [The CREMA Collaboration].} 
\newblock The helion charge radius from laser spectroscopy of muonic helium-3
  ions. \emph{ {arXiv}} {\bf 2023},  {arXiv:2305.11679}.
\newblock {\url{https://doi.org/10.48550/arXiv.2305.11679}}.

\bibitem[Krauth et~al.(2021)Krauth, Schuhmann, Ahmed, Amaro, Amaro, Biraben,
  Chen, Covita, Dax, Diepold, Fernandes, Franke, Galtier, Gouvea, Götzfried,
  Graf, Hänsch, Hartmann, Hildebrandt, Indelicato, Julien, Kirch, Knecht, Liu,
  Machado, Monteiro, Mulhauser, Naar, Nebel, Nez, dos Santos, Santos, Szabo,
  Taqqu, Veloso, Vogelsang, Voss, Weichelt, Pohl, Antognini, and
  Kottmann]{2021-Krauth}
Krauth, J.J.; Schuhmann, K.; Ahmed, M.A.; Amaro, F.D.; Amaro, P.; Biraben, F.;
  Chen, T.L.; Covita, D.S.; Dax, A.J.; Diepold, M.;~et~al.
\newblock Measuring the $\alpha$-particle charge radius with muonic helium-4
  ions.
\newblock {\em Nature} {\bf 2021}, {\em 589},~527--531.
\newblock {\url{https://doi.org/10.1038/s41586-021-03183-1}}.

\bibitem[Suelzle et~al.(1967)Suelzle, Yearian, and Crannell]{1967-LiScat}
Suelzle, L.R.; Yearian, M.R.; Crannell, H.
\newblock Elastic electron scattering from ${\mathrm{Li}}^{6}$ and
  ${\mathrm{Li}}^{7}$.
\newblock {\em Phys. Rev.} {\bf 1967}, {\em 162},~992--1005.
\newblock {\url{https://doi.org/10.1103/PhysRev.162.992}}.

\bibitem[Bumiller et~al.(1972)Bumiller, Buskirk, Dyer, and Monson]{1972-LiScat}
Bumiller, F.A.; Buskirk, F.R.; Dyer, J.N.; Monson, W.A.
\newblock Elastic electron scattering from $^{6}\mathrm{Li}$ and
  $^{7}\mathrm{Li}$ at low momentum transfer.
\newblock {\em Phys. Rev. C} {\bf 1972}, {\em 5},~391--395.
\newblock {\url{https://doi.org/10.1103/PhysRevC.5.391}}.

\bibitem[Li et~al.(1971)Li, Sick, Whitney, and Yearian]{1971-LiScat}
Li, G.; Sick, I.; Whitney, R.; Yearian, M.
\newblock High-energy electron scattering from  {$^6$}Li.
\newblock {\em Nucl. Phys. A} {\bf 1971}, {\em 162},~583--592.
\newblock {\url{https://doi.org/10.1016/0375-9474(71)90257-0}}.

\bibitem[N\"ortersh\"auser et~al.(2011{\natexlab{a}})N\"ortersh\"auser, Neff,
  S\'anchez, and Sick]{2011-Li}
N\"ortersh\"auser, W.; Neff, T.; S\'anchez, R.; Sick, I.
\newblock Charge radii and ground state structure of lithium isotopes:
  Experiment and theory reexamined.
\newblock {\em Phys. Rev. C} {\bf 2011}, {\em 84},~024307.
\newblock {\url{https://doi.org/10.1103/PhysRevC.84.024307}}.

\bibitem[N\"ortersh\"auser et~al.(2011{\natexlab{b}})N\"ortersh\"auser,
  S\'anchez, Ewald, Dax, Behr, Bricault, Bushaw, Dilling, Dombsky, Drake,
  G\"otte, Kluge, K\"uhl, Lassen, Levy, Pachucki, Pearson, Puchalski,
  Wojtaszek, Yan, and Zimmermann]{2011-LiIS}
N\"ortersh\"auser, W.; S\'anchez, R.; Ewald, G.; Dax, A.; Behr, J.; Bricault,
  P.; Bushaw, B.A.; Dilling, J.; Dombsky, M.; Drake, G.W.F.;~et~al.
\newblock Isotope-shift measurements of stable and short-lived lithium isotopes
  for nuclear-charge-radii determination.
\newblock {\em Phys. Rev. A} {\bf 2011}, {\em 83},~012516.
\newblock {\url{https://doi.org/10.1103/PhysRevA.83.012516}}.

\bibitem[Jansen et~al.(1972)Jansen, Peerdeman, and {De Vries}]{1972-Be}
Jansen, J. {A.}; Peerdeman, R. {T.}; {De Vries}, C.
\newblock Nuclear charge radii of  {$^{12}$}C and  {$^9$}Be.
\newblock {\em Nucl. Phys. A} {\bf 1972}, {\em 188},~337--352.
\newblock {\url{https://doi.org/10.1016/0375-9474(72)90062-0}}.

\bibitem[Maa\ss{} et~al.(2019)Maa\ss{}, H\"uther, K\"onig, Kr\"amer, Krause,
  Lovato, M\"uller, Pachucki, Puchalski, Roth, S\'anchez, Sommer, Wiringa, and
  N\"ortersh\"auser]{2019-B}
Maa\ss{}, B.; H\"uther, T.; K\"onig, K.; Kr\"amer, J.; Krause, J.; Lovato, A.;
  M\"uller, P.; Pachucki, K.; Puchalski, M.; Roth, R.;~et~al.
\newblock Nuclear charge radii of $^{10,11}\mathrm{B}$.
\newblock {\em Phys. Rev. Lett.} {\bf 2019}, {\em 122},~182501.
\newblock {\url{https://doi.org/10.1103/PhysRevLett.122.182501}}.

\bibitem[Barnett et~al.(1980)Barnett, Gyles, Johnson, Erdman, Johnstone,
  Kraushaar, Lepp, Masterson, Rost, Gill, Thomas, Alster, Navon, and
  Landau]{1980-B}
Barnett, B.; Gyles, W.; Johnson, R.; Erdman, K.; Johnstone, J.; Kraushaar, J.;
  Lepp, S.; Masterson, T.; Rost, E.; Gill, D.;~et~al.
\newblock Proton radii determinations from the ratio of $\pi^+$ elastic
  scattering from  {$^{11}$}B and  {$^{12}$}C.
\newblock {\em Phys. Lett. B} {\bf 1980}, {\em 97},~45--49.
\newblock {\url{https://doi.org/10.1016/0370-2693(80)90543-2}}.

\bibitem[Ruckstuhl et~al.(1984)Ruckstuhl, Aas, Beer, Beltrami, Bos, Goudsmit,
  Leisi, Strassner, Vacchi, De~Boer, Kiebele, and Weber]{1984-Ruckstuhl}
Ruckstuhl, W.; Aas, B.; Beer, W.; Beltrami, I.; Bos, K.; Goudsmit, P.F.A.;
  Leisi, H.J.; Strassner, G.; Vacchi, A.; De~Boer, F.W.N.;~et~al.
\newblock Precision measurement of the 2p-1s transition in muonic  {$^{12}$}C: Search
  for new muon-nucleon interactions or accurate determination of the rms
  nuclear charge radius.
\newblock {\em Nucl. Phys. A} {\bf 1984}, {\em 430},~685--712.
\newblock {\url{https://doi.org/10.1016/0375-9474(84)90101-5}}.

\bibitem[{de Boer} et~al.(1985){de Boer}, Aas, Baertschi, Beer, Beltrami, Bos,
  Goudsmit, Kiebele, Jeckelmann, Leisi, Ruckstuhl, Strassner, Vacchi, and
  Weber]{1985-13C}
{de Boer}, F. {W.N.;} Aas, B.; Baertschi, P.; Beer, W.; Beltrami, I.; Bos, K.;
  Goudsmit, P. {F.A.}; Kiebele, U.; Jeckelmann, B.; Leisi, H. {J.};~et~al.
\newblock Precision measurement of the 2p-1s transition wavelength in muonic
   {$^{13}$}C.
\newblock {\em Nucl. Phys. A} {\bf 1985}, {\em 444},~589--596.
\newblock {\url{https://doi.org/10.1016/0375-9474(85)90106-X}}.

\bibitem[Epelbaum et~al.(2009)Epelbaum, Hammer, and Mei\ss{}ner]{2009-Epelbaum}
Epelbaum, E.; Hammer, H. {-}W.; Mei\ss{}ner, U. {-}G.
\newblock Modern theory of nuclear forces.
\newblock {\em Rev. Mod. Phys.} {\bf 2009}, {\em 81},~1773--1825.
\newblock {\url{https://doi.org/10.1103/RevModPhys.81.1773}}.

\bibitem[Hergert(2020)]{2020-Hergert}
Hergert, H.
\newblock A Guided tour of ab initio nuclear many-body theory.
\newblock {\em Front. Phys.} {\bf 2020}, {\em 8}, 379.
\newblock {\url{https://doi.org/10.3389/fphy.2020.00379}}.

\bibitem[Dohet-Eraly et~al.(2016)Dohet-Eraly, Navrátil, Quaglioni, Horiuchi,
  Hupin, and Raimondi]{2016-NCSMC}
Dohet-Eraly, J.; Navr\'atil, P.; Quaglioni, S.; Horiuchi, W.; Hupin, G.;
  Raimondi, F.
\newblock  {$^3$He($\alpha$,$\gamma$)$\,^7$Be} and  {$^3$H($\alpha$,$\gamma$)$\,^7$Li} astrophysical
   {$S$} factors from the no-core shell model with continuum.
\newblock {\em Phys. Lett. B} {\bf 2016}, {\em 757},~430--436.
\newblock
  {\url{https://doi.org/10.1016/j.physletb.2016.04.021}}.

\bibitem[Navr{\'{a}}til et~al.(2016)Navr{\'{a}}til, Quaglioni, Hupin,
  Romero-Redondo, and Calci]{2016-NAv}
Navr{\'{a}}til, P.; Quaglioni, S.; Hupin, G.; Romero-Redondo, C.; Calci, A.
\newblock Unified ab initio approaches to nuclear structure and reactions.
\newblock {\em Phys. Scr.} {\bf 2016}, {\em 91},~053002.
\newblock {\url{https://doi.org/10.1088/0031-8949/91/5/053002}}.

\bibitem[Vorabbi et~al.(2019)Vorabbi, Navr\'atil, Quaglioni, and
  Hupin]{2019-Vorrabi}
Vorabbi, M.; Navr\'atil, P.; Quaglioni, S.; Hupin, G.
\newblock $^{7}\mathrm{Be}$ and $^{7}\mathrm{Li}$ nuclei within the no-core
  shell model with continuum.
\newblock {\em Phys. Rev. C} {\bf 2019}, {\em 100},~024304.
\newblock {\url{https://doi.org/10.1103/PhysRevC.100.024304}}.

\bibitem[Caprio et~al.(2022)Caprio, Fasano, and Maris]{2022-Caprio}
Caprio, M.A.; Fasano, P.J.; Maris, P.
\newblock Robust ab initio prediction of nuclear electric quadrupole
  observables by scaling to the charge radius.
\newblock {\em Phys. Rev. C} {\bf 2022}, {\em 105},~L061302.
\newblock {\url{https://doi.org/10.1103/PhysRevC.105.L061302}}.

\bibitem[Ewald et~al.(2004)Ewald, N\"ortersh\"auser, Dax, G\"otte, Kirchner,
  Kluge, K\"uhl, Sanchez, Wojtaszek, Bushaw, Drake, Yan, and
  Zimmermann]{2004-Li89}
Ewald, G.; N\"ortersh\"auser, W.; Dax, A.; G\"otte, S.; Kirchner, R.; Kluge,
  H. {-}J.; K\"uhl, T.; Sanchez, R.; Wojtaszek, A.; Bushaw, B.A.;~et~al.
\newblock Nuclear charge radii of $^{8,9}\mathrm{L}\mathrm{i}$ determined by
  laser spectroscopy.
\newblock {\em Phys. Rev. Lett.} {\bf 2004}, {\em 93},~113002.
\newblock {\url{https://doi.org/10.1103/PhysRevLett.93.113002}}.

\bibitem[S\'anchez et~al.(2006)S\'anchez, N\"ortersh\"auser, Ewald, Albers,
  Behr, Bricault, Bushaw, Dax, Dilling, Dombsky, Drake, G\"otte, Kirchner,
  Kluge, K\"uhl, Lassen, Levy, Pearson, Prime, Ryjkov, Wojtaszek, Yan, and
  Zimmermann]{2006-Li911}
S\'anchez, R.; N\"ortersh\"auser, W.; Ewald, G.; Albers, D.; Behr, J.;
  Bricault, P.; Bushaw, B.A.; Dax, A.; Dilling, J.; Dombsky, M.;~et~al.
\newblock Nuclear charge radii of $^{9,11}\mathrm{Li}$: The influence of halo
  neutrons.
\newblock {\em Phys. Rev. Lett.} {\bf 2006}, {\em 96},~033002.
\newblock {\url{https://doi.org/10.1103/PhysRevLett.96.033002}}.

\bibitem[N\"ortersh\"auser et~al.(2009)]{2009-Be}
N\"ortersh\"auser, W.; Tiedemann, D.; \v{Z}\'akov\'a, M.; Andjelkovic, Z.; Blaum, K.; Bissell, M.L.; Cazan, R.; Drake, G.W.F.; Geppert, C.; Kowalska, 
M.;~et~al.
\newblock Nuclear charge radii of $^{7,9,10}\mathrm{Be}$ and the one-neutron
  halo nucleus $^{11}\mathrm{Be}$.
\newblock {\em Phys. Rev. Lett.} {\bf 2009}, {\em 102},~062503.
\newblock {\url{https://doi.org/10.1103/PhysRevLett.102.062503}}.

\bibitem[Brown(2017)]{2017-Brown}
Brown, B.A.
\newblock Mirror charge radii and the neutron equation of state.
\newblock {\em Phys. Rev. Lett.} {\bf 2017}, {\em 119},~122502.
\newblock {\url{https://doi.org/10.1103/PhysRevLett.119.122502}}.

\bibitem[Brown et~al.(2020)Brown, Minamisono, Piekarewicz, Hergert, Garand,
  Klose, K\"onig, Lantis, Liu, Maa\ss{}, Miller, N\"ortersh\"auser, Pineda,
  Powel, Rossi, Sommer, Sumithrarachchi, Teigelh\"ofer, Watkins, and
  Wirth]{2020-ArCa}
Brown, B.A.; Minamisono, K.; Piekarewicz, J.; Hergert, H.; Garand, D.; Klose,
  A.; K\"onig, K.; Lantis, J.D.; Liu, Y.; Maa\ss{}, B.;~et~al.
\newblock Implications of the $^{36}\mathrm{Ca}$--
$^{36}\mathrm{S}$
  and $^{38}\mathrm{Ca}$--
$^{38}\mathrm{Ar}$ difference in mirror
  charge radii on the neutron matter equation of state.
\newblock {\em Phys. Rev. Res.} {\bf 2020}, {\em 2},~022035.
\newblock {\url{https://doi.org/10.1103/PhysRevResearch.2.022035}}.

\bibitem[Pineda et~al.(2021)Pineda, K\"onig, Rossi, Brown, Incorvati, Lantis,
  Minamisono, N\"ortersh\"auser, Piekarewicz, Powel, and Sommer]{2021-Pineda}
Pineda, S.V.; K\"onig, K.; Rossi, D.M.; Brown, B.A.; Incorvati, A.; Lantis, J.;
  Minamisono, K.; N\"ortersh\"auser, W.; Piekarewicz, J.; Powel, R.;~et~al.
\newblock Charge radius of neutron-deficient $^{54}\mathrm{Ni}$ and symmetry
  energy constraints using the difference in mirror pair charge radii.
\newblock {\em Phys. Rev. Lett.} {\bf 2021}, {\em 127},~182503.
\newblock {\url{https://doi.org/10.1103/PhysRevLett.127.182503}}.

\bibitem[Naito et~al.(2022)Naito, Roca-Maza, Col\`o, Liang, and
  Sagawa]{2022-ISO}
Naito, T.; Roca-Maza, X.; Col\`o, G.; Liang, H.; Sagawa, H.
\newblock Isospin symmetry breaking in the charge radius difference of mirror
  nuclei.
\newblock {\em Phys. Rev. C} {\bf 2022}, {\em 106},~L061306.
\newblock {\url{https://doi.org/10.1103/PhysRevC.106.L061306}}.

\bibitem[Reinhard and Nazarewicz(2022)]{2022-info}
Reinhard, P.G.; Nazarewicz, W.
\newblock Information content of the differences in the charge radii of mirror
  nuclei.
\newblock {\em Phys. Rev. C} {\bf 2022}, {\em 105},~L021301.
\newblock {\url{https://doi.org/10.1103/PhysRevC.105.L021301}}.

\bibitem[Bano et~al.(2023)Bano, Pattnaik, Centelles, Vi\~nas, and
  Routray]{2023-MirrStellar}
Bano, P.; Pattnaik, S.P.; Centelles, M.; Vi\~nas, X.; Routray, T.R.
\newblock Correlations between charge radii differences of mirror nuclei and
  stellar observables.
\newblock {\em Phys. Rev. C} {\bf 2023}, {\em 108},~015802.
\newblock {\url{https://doi.org/10.1103/PhysRevC.108.015802}}.

\bibitem[Yang and Piekarewicz(2018)]{2018-Mirr}
Yang, J.; Piekarewicz, J.
\newblock Difference in proton radii of mirror nuclei as a possible surrogate
  for the neutron skin.
\newblock {\em Phys. Rev. C} {\bf 2018}, {\em 97},~014314.
\newblock {\url{https://doi.org/10.1103/PhysRevC.97.014314}}.

\bibitem[Gaidarov et~al.(2020)Gaidarov, Moumene, Antonov, Kadrev, Sarriguren,
  and {Moya de Guerra}]{2020-Mirr}
Gaidarov, M.; Moumene, I.; Antonov, A.; Kadrev, D.; Sarriguren, P.; {Moya de
  Guerra}, E.
\newblock Proton and neutron skins and symmetry energy of mirror nuclei.
\newblock {\em Nucl. Phys. A} {\bf 2020}, {\em 1004},~122061.
\newblock
  {\url{https://doi.org/10.1016/j.nuclphysa.2020.122061}}.

\bibitem[Novario et~al.(2023)Novario, Lonardoni, Gandolfi, and
  Hagen]{2023-Mirr}
Novario, S.J.; Lonardoni, D.; Gandolfi, S.; Hagen, G.
\newblock Trends of neutron skins and radii of mirror nuclei from first
  principles.
\newblock {\em Phys. Rev. Lett.} {\bf 2023}, {\em 130},~032501.
\newblock {\url{https://doi.org/10.1103/PhysRevLett.130.032501}}.

\bibitem[Maa{\ss} et~al.(2017)]{2017-B8}
Maa\ss{}, B.; M\"uller, P.; N\"ortersh\"auser, W.; Clark, J.; Gorges, C.; Kaufmann, S.; K\"onig, K.; Kr\"amer, J.; Levand, A.; Orford, R.;~et~al.
\newblock Towards laser spectroscopy of the proton-halo candidate boron-8.
\newblock {\em Hyperfine Interact.} {\bf 2017}, {\em 238},~25.
\newblock {\url{https://doi.org/10.1007/s10751-017-1399-5}}.

\bibitem[Yerokhin and Pachucki(2010)]{2010-Yerokhin}
Yerokhin, V.A.; Pachucki, K.
\newblock Theoretical energies of low-lying states of light helium-like ions.
\newblock {\em Phys. Rev. A} {\bf 2010}, {\em 81},~022507.
\newblock {\url{https://doi.org/10.1103/PhysRevA.81.022507}}.

\bibitem[Yerokhin et~al.(2022)Yerokhin, Patk\'o\ifmmode~\check{s}\else
  \v{s}\fi{}, and Pachucki]{2022-Yerokhin}
Yerokhin, V.A.;  {Patk\'o\v{s}, V.; 
 Pachucki, K.}
\newblock QED calculations of energy levels of heliumlike ions with
  $5\ensuremath{\le}Z\ensuremath{\le}30$.
\newblock {\em Phys. Rev. A} {\bf 2022}, {\em 106},~022815.
\newblock {\url{https://doi.org/10.1103/PhysRevA.106.022815}}.

\bibitem[Yerokhin et~al.(2023)Yerokhin, Patk\'o\ifmmode~\check{s}\else
  \v{s}\fi{}, and Pachucki]{2023-HLI}
Yerokhin, V.A.; 
  {Patk\'o\v{s}, V.; 
 Pachucki,} K.
\newblock QED $m{\ensuremath{\alpha}}^{7}$ effects for triplet states of
  heliumlike ions.
\newblock {\em Phys. Rev. A} {\bf 2023}, {\em 107},~012810.
\newblock {\url{https://doi.org/10.1103/PhysRevA.107.012810}}.

\bibitem[Riis et~al.(1994)Riis, Sinclair, Poulsen, Drake, Rowley, and
  Levick]{1994-Li}
Riis, E.; Sinclair, A.G.; Poulsen, O.; Drake, G.W.F.; Rowley, W.R.C.; Levick,
  A.P.
\newblock Lamb shifts and hyperfine structure in $^{6}\mathrm{Li}^{+}$ and
  $^{7}\mathrm{Li}^{+}$: Theory and experiment.
\newblock {\em Phys. Rev. A} {\bf 1994}, {\em 49},~207--220.
\newblock {\url{https://doi.org/10.1103/PhysRevA.49.207}}.

\bibitem[Guan et~al.(2020)Guan, Chen, Qi, Liang, Sun, Zhou, Huang, Zhang,
  Zhong, Yan, Drake, Shi, and Gao]{2020-LiOngoing}
Guan, H.; Chen, S.; Qi, X. {-}Q.; Liang, S.; Sun, W.; Zhou, P.; Huang, Y.; Zhang,
  P. {-}P.; Zhong, Z. {-}X.; Yan, Z. {-}C.;~et~al.
\newblock Probing atomic and nuclear properties with precision spectroscopy of
  fine and hyperfine structures in the $^{7}{\mathrm{Li}}^{+}$ ion.
\newblock {\em Phys. Rev. A} {\bf 2020}, {\em 102},~030801.
\newblock {\url{https://doi.org/10.1103/PhysRevA.102.030801}}.

\bibitem[Clausen et~al.(2021)Clausen, Jansen, Scheidegger, Agner, Schmutz, and
  Merkt]{2021-He}
Clausen, G.; Jansen, P.; Scheidegger, S.; Agner, J.A.; Schmutz, H.; Merkt, F.
\newblock Ionization energy of the metastable $2\text{ }^{1}{\mathrm{S}}_{0}$
  state of $^{4}\mathrm{He}$ from Rydberg-series extrapolation.
\newblock {\em Phys. Rev. Lett.} {\bf 2021}, {\em 127},~093001.
\newblock {\url{https://doi.org/10.1103/PhysRevLett.127.093001}}.

\bibitem[Clausen et~al.(2023)Clausen, Scheidegger, Agner, Schmutz, and
  Merkt]{2023-He}
Clausen, G.; Scheidegger, S.; Agner, J.A.; Schmutz, H.; Merkt, F.
\newblock Imaging-assisted single-photon Doppler-free laser spectroscopy and
  the ionization energy of metastable triplet helium.
\newblock {\em Phys. Rev. Lett.} {\bf 2023}, {\em 131},~103001.
\newblock {\url{https://doi.org/10.1103/PhysRevLett.131.103001}}.

\bibitem[Maa{\ss}(2020)]{2020-BHLI}
Maa{\ss}, B.
\newblock Laser Spectroscopy of the Boron Isotopic Chain.
\newblock Ph.D. Thesis, TU Darmstadt,  {Darmstadt, Germany,}  2020.
\newblock {\url{https://doi.org/10.25534/tuprints-00011484}}.

\bibitem[König et~al.(2020)]{2020-COALA}
K\"onig, K.;   {Kr\"amer, K.;
 Geppert, C.;
Imgram, P.;
 Maa{\ss}, B.;
Ratajczyk, T.;
N\"ortersh\"auser, W.}  
\newblock A new Collinear Apparatus for Laser Spectroscopy and Applied Science
  (COALA).
\newblock {\em Rev. Sci. Instrum.} {\bf 2020}, {\em 91},~081301.
\newblock {\url{https://doi.org/10.1063/5.0010903}}.

\bibitem[Mohr et~al.(2023)Mohr, Buß, Andelkovic, Hannen, Horst, Imgram,
  König, Maaß, Nörtershäuser, Rausch, Sánchez, and Weinheimer]{2023-HLIB}
Mohr, K.; Bu{\ss}, A.; Andelkovic, Z.; Hannen, V.; Horst, M.; Imgram, P.; K\"onig,
  K.; Maa{\ss}, B.; N\"ortersh\"auser, W.; Rausch, S.;~et~al.
\newblock Collinear laser spectroscopy of helium-like $^{11}$B$^{3+}$.
\newblock {\em Atoms} {\bf 2023}, {\em 11}, 11.
\newblock {\url{https://doi.org/10.3390/atoms11010011}}.

\bibitem[Imgram(2023)]{2023-HeC}
Imgram, P.
\newblock High-Precision Laser Spectroscopy of Helium-like Carbon
  $^{12}$C$^{4+}$.  {Ph.D. Thesis, TU Darmstadt,} 
  {Darmstadt, Germany,} 
  2023.
\newblock {\url{https://doi.org/10.26083/tuprints-00023082}}.

\bibitem[Imgram et~al.(2023{\natexlab{a}})Imgram, K\"onig, Maa\ss{}, M\"uller,
  and N\"ortersh\"auser]{2023-HLIPRA}
Imgram, P.; K\"onig, K.; Maa\ss{}, B.; M\"uller, P.; N\"ortersh\"auser, W.
\newblock Collinear laser spectroscopy of highly charged ions produced with an
  electron-beam ion source.
\newblock {\em Phys. Rev. A} {\bf 2023}, {\em 108},~062809.
\newblock {\url{https://doi.org/10.1103/PhysRevA.108.062809}}.

\bibitem[Imgram et~al.(2023{\natexlab{b}})Imgram, K\"onig, Maa\ss{}, M\"uller,
  and N\"ortersh\"auser]{2023-HLIPRL}
Imgram, P.; K\"onig, K.; Maa\ss{}, B.; M\"uller, P.; N\"ortersh\"auser, W.
\newblock Collinear laser spectroscopy of $2\text{
  }^{3}{S}_{1}\ensuremath{\rightarrow}2\text{ }^{3}{P}_{J}$ transitions in
  helium-like ${^{12}\mathrm{C}}^{4+}$.
\newblock {\em Phys. Rev. Lett.} {\bf 2023}, {\em 131},~243001.
\newblock {\url{https://doi.org/10.1103/PhysRevLett.131.243001}}.

\bibitem[Sailer et~al.(2022)Sailer, Debierre, Harman, Hei{\ss}e, K{\"o}nig,
  Morgner, Tu, Volotka, Keitel, Blaum, et~al.]{2022-Sailer}
Sailer, T.; Debierre, V.; Harman, Z.; Hei{\ss}e, F.; K{\"o}nig, C.; Morgner,
  J.; Tu, B.; Volotka, A.V.; Keitel, C.H.; Blaum, K.; Sturm, S. 
\newblock Measurement of the bound-electron $g$-factor difference in coupled
  ions.
\newblock {\em Nature} {\bf 2022}, {\em 606},~479--483.
\newblock {\url{https://doi.org/10.1038\%2Fs41586-022-04807-w}}.

\bibitem[Schmidt et~al.(2018)Schmidt, Willig, Haack, Horn, Adamczak, Ahmed,
  Amaro, Amaro, Biraben, Carvalho, Chen, Fernandes, Graf, Guerra, Haensch,
  Hildebrandt, Huang, Indelicato, Julien, and Pohl]{2018-LaserNext}
Schmidt, S.; Willig, M.; Haack, J.; Horn, R.; Adamczak, A.;  {Abdou} Ahmed, M.; Amaro,
  F. {D.;} Amaro, P.; Biraben, F.; Carvalho, P.;~et~al.
\newblock The next generation of laser spectroscopy experiments using light
  muonic atoms.
\newblock {\em J. Phys. Conf. Ser.} {\bf 2018}, {\em
  1138},~012010.
\newblock {\url{https://doi.org/10.1088/1742-6596/1138/1/012010}}.

\bibitem[Yerokhin(2008)]{2008-Li}
Yerokhin, V.A.
\newblock Hyperfine structure of Li and ${\mathrm{Be}}^{+}$.
\newblock {\em Phys. Rev. A} {\bf 2008}, {\em 78},~012513.
\newblock {\url{https://doi.org/10.1103/PhysRevA.78.012513}}.

\bibitem[Sun et~al.(2023)Sun, Zhang, Zhou, Chen, Zhou, Huang, Qi, Yan, Shi,
  Drake, Zhong, Guan, and Gao]{2023-LiHFS}
Sun, W.; Zhang,  {P.-P.;} Zhou,  {P.-p.}; Chen,  {S.-l.}; Zhou,  {Z.-q.}; Huang, Y.; Qi,  {X.-Q.};
  Yan,  {Z.-C.}; Shi,  {T.-Y.}; Drake, G.W.F.;~et~al.
\newblock Measurement of hyperfine structure and the Zemach radius in
  $^{6}{\mathrm{Li}}^{+}$ using optical Ramsey technique.
\newblock {\em Phys. Rev. Lett.} {\bf 2023}, {\em 131},~103002.
\newblock {\url{https://doi.org/10.1103/PhysRevLett.131.103002}}.

\bibitem[Pachucki et~al.(2023)Pachucki, Patk\'o\ifmmode~\check{s}\else
  \v{s}\fi{}, and Yerokhin]{2023-LiHFSTh}
Pachucki, K.; 
 Patk\'o\v{s}, V.; 
 Yerokhin, V.A.  
\newblock Hyperfine splitting in $^{6,7}\mathrm{Li}^{+}$.
\newblock {\em Phys. Rev. A} {\bf 2023}, {\em 108},~052802.
\newblock {\url{https://doi.org/10.1103/PhysRevA.108.052802}}.

\bibitem[202(2023)]{2023-LiMoments}
Pachucki, K.; 
 Patk\'o\v{s}, V.;   
 Yerokhin, V.A.
Accurate determination of $^{6,7}\mathrm{Li}$ nuclear magnetic moments.
\newblock {\em Phys. Lett. B} {\bf 2023}, {\em 846},~138189.
\newblock
  {\url{https://doi.org/10.1016/j.physletb.2023.138189}}.

\bibitem[Krutov et~al.(2016)Krutov, Martynenko, Martynenko, and
  Sukhorukova]{2016-LS_Light}
Krutov, A.A.; Martynenko, A.P.; Martynenko, F.A.; Sukhorukova, O.S.
\newblock Lamb shift in muonic ions of lithium, beryllium, and boron.
\newblock {\em Phys. Rev. A} {\bf 2016}, {\em 94},~062505.
\newblock {\url{https://doi.org/10.1103/PhysRevA.94.062505}}.

\bibitem[Dorokhov et~al.(2021)Dorokhov, Faustov, Martynenko, and
  Martynenko]{2021-1S-2S}
Dorokhov, A.E.; Faustov, R.N.; Martynenko, A.P.; Martynenko, F.A.
\newblock Precision physics of muonic ions of lithium, beryllium and boron.
\newblock {\em Int. J. Mod. Phys. A} {\bf 2021}, {\em
  36},~2150022.
\newblock {\url{https://doi.org/10.1142/S0217751X21500226}}.

\bibitem[Li~Muli et~al.(2020)Li~Muli, Poggialini, and Bacca]{Muli2020}
Li~Muli, S. {S.}; Poggialini, A.; Bacca, S.
\newblock {Muonic lithium atoms: Nuclear structure corrections to the Lamb
  shift}.
\newblock {\em SciPost Phys. Proc.} {\bf 2020}, {\it  {3}},  {028.} 
\newblock {\url{https://doi.org/10.21468/SciPostPhysProc.3.028}}.

\bibitem[Valuev et~al.(2022)Valuev, Col\`o, Roca-Maza, Keitel, and
  Oreshkina]{2022-NP}
Valuev, I.A.; Col\`o, G.; Roca-Maza, X.; Keitel, C.H.; Oreshkina, N.S.
\newblock Evidence against nuclear polarization as source of fine-structure
  anomalies in muonic atoms.
\newblock {\em Phys. Rev. Lett.} {\bf 2022}, {\em 128},~203001.
\newblock {\url{https://doi.org/10.1103/PhysRevLett.128.203001}}.

\bibitem[Barrett et~al.(2013)Barrett, Navr\'{a}til, and Vary]{2013-NCSM}
Barrett, B.R.; Navr\'{a}til, P.; Vary, J.P.
\newblock {Ab initio no core shell model}.
\newblock {\em Prog. Part. Nucl. Phys.} {\bf 2013}, {\em 69},~131--181.
\newblock {\url{https://doi.org/10.1016/j.ppnp.2012.10.003}}.

\bibitem[Lanczos(1950)]{1950-Lanc}
Lanczos, C.
\newblock An iteration method for the solution of the eigenvalue problem of
  linear differential and integral operators.
\newblock {\em J. Res. Natl. Bur. Stand.} {\bf 1950}, {\em 45},~255--282.
\newblock {\url{https://doi.org/10.6028/jres.045.026}}.

\bibitem[Haydock(1974)]{1974-Hay}
Haydock, R.
\newblock The inverse of a linear operator.
\newblock {\em J. Phys. Math. Nucl. Gen.} {\bf
  1974}, {\em 7},~2120--2124.
\newblock {\url{https://doi.org/10.1088/0305-4470/7/17/006}}.

\bibitem[Reuter et~al.(1982)Reuter, Fricke, Merle, and Miska]{1982-ScatC}
Reuter, W.; Fricke, G.; Merle, K.; Miska, H.
\newblock Nuclear charge distribution and rms radius of $^{12}\mathrm{C}$ from
  absolute elastic electron scattering measurements.
\newblock {\em Phys. Rev. C} {\bf 1982}, {\em 26},~806--818.
\newblock {\url{https://doi.org/10.1103/PhysRevC.26.806}}.

\bibitem[Schneuwly and Vogel(1980)]{1980-screen}
Schneuwly, H.; Vogel, P.
\newblock Electronic $K$ x-ray energies in heavy muonic atoms.
\newblock {\em Phys. Rev. A} {\bf 1980}, {\em 22},~2081--2087.
\newblock {\url{https://doi.org/10.1103/PhysRevA.22.2081}}.

\bibitem[199(1994)]{1994-screen}
 {Simons, L.M.;  
Abbot, D.;  
Bach, B.;  
Bacher, R.; 
Badertscher, A.; 
Bl\"um, P.;  
DeCecco, P.;  
Eades, J.; 
Egger, J.;  
Elsener, K.; et al.}   
Exotic atoms and their electron shell.
\newblock {\em Nucl. Instrum. Meth. Phys. Res. 
B Beam Interact. Mater. Atoms} {\bf 1994}, {\em 87},~293--300.
\newblock {\url{https://doi.org/10.1016/0168-583X(94)95275-2}}.

\bibitem[Desclaux(1975)]{1975-Des}
Desclaux, J. {P.} 
\newblock A multiconfiguration relativistic DIRAC-FOCK program.
\newblock {\em Comput. Phys. Commun.} {\bf 1975}, {\em 9},~31--45.
\newblock {\url{https://doi.org/10.1016/0010-4655(75)90054-5}}.

\bibitem[Mallow et~al.(1978)Mallow, Desclaux, and Freeman]{1978-MDF}
Mallow, J.V.; Desclaux, J.P.; Freeman, A.J.
\newblock \relax{Dirac-Fock method for muonic atoms : Transitions energies,
  wave functions, and charge densities}.
\newblock {\em Phys. Rev. A} {\bf 1978}, {\em 17},~1804 {--1809.}
\newblock {\url{https://doi.org/10.1103/PhysRevA.17.1804}}.

\bibitem[Santos et~al.(2005)Santos, Parente, Boucard, Indelicato, and
  Desclaux]{2005-SPBI}
Santos, J.P.; Parente, F.; Boucard, S.; Indelicato, P.; Desclaux, J.P.
\newblock \relax{X-ray energies of circular transitions and electron screening
  in kaonic atoms}.
\newblock {\em Phys. Rev. A} {\bf 2005}, {\em 71},~032501.
\newblock {\url{https://doi.org/10.1103/PhysRevA.71.032501}}.

\bibitem[Trassinelli and Indelicato(2007)]{2007-TAI}
Trassinelli, M.; Indelicato, P.
\newblock \relax{Relativistic calculations of pionic and kaonic atoms'
  hyperfine structure }.
\newblock {\em Phys. Rev. A} {\bf 2007}, {\em 76},~012510.
\newblock {\url{https://doi.org/10.1103/PhysRevA.76.012510}}.

\bibitem[Indelicato(2013)]{2013-Indel}
Indelicato, P.
\newblock Nonperturbative evaluation of some QED contributions to the muonic
  hydrogen $n=2$ Lamb shift and hyperfine structure.
\newblock {\em Phys. Rev. A} {\bf 2013}, {\em 87},~022501.
\newblock {\url{https://doi.org/10.1103/PhysRevA.87.022501}}.

\bibitem[{De Wit} et~al.(1966){De Wit}, Backenstoss, Daum, Sens, and
  Acker]{1966-def}
{De Wit}, S. {A.}; Backenstoss, G.; Daum, C.; Sens, J. {C.}; Acker, H. {L.}
\newblock Measurement and analysis of muonic X-ray spectra in deformed nuclei.
\newblock {\em Nucl. Phys.} {\bf 1966}, {\em 87},~657--702.
\newblock {\url{https://doi.org/10.1016/0029-5582(67)90003-X}}.

\bibitem[Pieper and Greiner(1967)]{1967-def}
Pieper, W.; Greiner, W.
\newblock The influence of nuclear dynamics on the x-ray spectrum of muonic
  atoms.
\newblock {\em Phys. Lett. B} {\bf 1967}, {\em 24},~377--380.
\newblock {\url{https://doi.org/10.1016/0370-2693(67)90295-X}}.

\bibitem[Foot et~al.(1982)Foot, Stacey, Stacey, Kloch, and
  Le{\'s}]{1982-ZnDeform}
Foot, C. {J.}; Stacey, D. {N.}; Stacey, V.; Kloch, R.; Le{\'s}, Z.
\newblock Isotope effects in the nuclear charge distribution in zinc.
\newblock {\it Proc. R. Soc.  {A} 
 Math. Phys.  {Engin.}  Sci.} {\bf 1982}, {\em 384},~205--216.
  {https://doi.org/10.1098/rspa.1982.0155} 

\bibitem[Sahoo and Ohayon(2023)]{2023-Zn}
Sahoo, B.K.; Ohayon, B.
\newblock All-optical differential radii in zinc.
\newblock {\em Phys. Rev. Res.} {\bf 2023}, {\em 5},~043142.
\newblock {\url{https://doi.org/10.1103/PhysRevResearch.5.043142}}.

\bibitem[Çavuşoğlu and Sikora(2023)]{2023-def}
\c{C}avu\c{s}o\v{g}lu, A.; Sikora, B.
\newblock Impact of the nuclear charge distribution on the g-factors and ground
  state energies of bound muons. \emph{ {arXiv}} \textbf{2023},  {arXiv:2311.16855}.
  \url{http://arxiv.org/abs/2311.16855}.

\bibitem[Paul et~al.(2021)Paul, Bian, Azuma, Okada, and
  Indelicato]{2021-Exotic}
Paul, N.; Bian, G.; Azuma, T.; Okada, S.; Indelicato, P.
\newblock Testing quantum electrodynamics with exotic atoms.
\newblock {\em Phys. Rev. Lett.} {\bf 2021}, {\em 126},~173001.
\newblock {\url{https://doi.org/10.1103/PhysRevLett.126.173001}}.

\bibitem[Indelicato(2019)]{2019-QEDHCI}
Indelicato, P.
\newblock {QED} tests with highly charged ions.
\newblock {\em J. Phys. B At. Mol. Opt. Phys.}
  {\bf 2019}, {\em 52},~232001.
\newblock {\url{https://doi.org/10.1088/1361-6455/ab42c9}}.

\bibitem[Fricke et~al.(2004)Fricke, Heilig, and Schopper]{2004-Fricke}
 {Fricke, G.; Heilig, K.}  
\newblock \emph{Nuclear Charge Radii}; 
 {Springer Nature Switzerland AG: Cham, Switzerland, } 
2004. 
\newblock {\url{https://doi.org/10.1007/b87879}}.

\bibitem[Simons(1999)]{1999-XRAY}
Simons, L.M.
\newblock X-ray spectroscopy at PSI.
\newblock {\em Hyperfine Interact.} {\bf 1999}, {\em 119},~281--290.
\newblock {\url{https://doi.org/10.1023/A:1012668316155}}.

\bibitem[Ruckstuhl et~al.(1982)Ruckstuhl, Aas, Beer, Beltrami, de~Boer, Bos,
  Goudsmit, Kiebele, Leisi, Strassner, Vacchi, and Weber]{1982-Ruckstuhl}
Ruckstuhl, W.; Aas, B.; Beer, W.; Beltrami, I.; de~Boer, F.W.N.; Bos, K.;
  Goudsmit, P.F.A.; Kiebele, U.; Leisi, H.J.; Strassner, G.;~et~al.
\newblock High-precision muonic X-ray measurement of the rms charge radius of
  $^{12}\mathrm{C}$ with a crystal spectrometer.
\newblock {\em Phys. Rev. Lett.} {\bf 1982}, {\em 49},~859--862.
\newblock {\url{https://doi.org/10.1103/PhysRevLett.49.859}}.

\bibitem[Fleischmann et~al.(2005)Fleischmann, Enss, and 
  Seidel]{2005-Fleischmann}
Fleischmann, A.; Enss, C.; Seidel, G. Metallic magnetic calorimeters.
\newblock In {\em Cryogenic Particle Detection}; Enss, C., Ed.; Springer: Berlin/Heidelberg, Germany, 2005; pp. 151--216.
\newblock {\url{https://doi.org/10.1007/10933596_4}}.

\bibitem[Unger et~al.(2023)Unger, Abeln, Cocolios, Eizenberg, Enss,
  Fleischmann, Gastaldo, Godinho, Heines, Hengstler, Indelicato, Jadhav,
  Kreuzberger, Kirch, Knecht, Machado, Ohayon, Paul, Pohl, von Schoeler,
  Vogiatzi, and Wauters]{other}
Unger, D.; Abeln, A.; Cocolios, T.E.; Eizenberg, O.; Enss, C.; Fleischmann, A.;
  Gastaldo, L.; Godinho, C.; Heines, M.; Hengstler, D.;~et~al.
\newblock MMC array to study X-ray transitions in muonic atoms. \emph{ {arXiv}} {\bf 2023},  {arXiv:2311.12014}.
\newblock {\url{https://doi.org/10.48550/arXiv.2311.12014}}.

\bibitem[Hengstler et~al.(2015)Hengstler, Keller, Schötz, Geist, Krantz,
  Kempf, Gastaldo, Fleischmann, Gassner, a, et~al.]{2015-MAXS}
Hengstler, D.; Keller, M.; Sch\"oltz, C.; Geist, J.; Krantz, M.; Kempf, S.;
  Gastaldo, L.; Fleischmann, A.; Gassner, T.; Weber, G.;~et~al.
\newblock Towards {FAIR}: First measurements of metallic magnetic calorimeters
  for high-resolution {X-ray} spectroscopy at {GSI}.
\newblock {\em Phys. Scr.} {\bf 2015}, {\em T166},~014054.
\newblock {\url{https://doi.org/10.1088/0031-8949/2015/t166/014054}}.

\bibitem[Unger et~al.(2021)Unger, Abeln, Enss, Fleischmann, Hengstler, Kempf,
  and Gastaldo]{2021-Unger}
Unger, D.; Abeln, A.; Enss, C.; Fleischmann, A.; Hengstler, D.; Kempf, S.;
  Gastaldo, L.
\newblock High-resolution for IAXO: MMC-based X-ray detectors.
\newblock {\em J. Instrum.} {\bf 2021}, {\em 16},~P06006.
\newblock {\url{https://doi.org/10.1088/1748-0221/16/06/P06006}}.

\bibitem[Wauters and Knecht(2021)]{2021-MuX}
Wauters, F.; Knecht, A.
\newblock {The muX project}.
\newblock {\em SciPost Phys. Proc.} {\bf 2021}, {\em 5},~022.
\newblock {\url{https://doi.org/10.21468/SciPostPhysProc.5.022}}.

\bibitem[Boyd et~al.(2020)Boyd, Kim, Hall, Cantor, and Friedrich]{2020-MMCLin}
Boyd, S.; Kim, G.B.; Hall, J.; Cantor, R.; Friedrich, S.
\newblock Metallic magnetic calorimeters for high-accuracy nuclear decay data.
\newblock {\em J. Low Temp. Phys.} {\bf 2020}, {\em
  199},~681--687.
\newblock {\url{https://doi.org/10.1007/s10909-020-02406-5}}.

\bibitem[Deslattes et~al.(2003)Deslattes, Kessler, Indelicato, de~Billy,
  Lindroth, and Anton]{2003-XRAY}
Deslattes, R.D.; Kessler, E.G.; Indelicato, P.; de~Billy, L.; Lindroth, E.;
  Anton, J.
\newblock X-ray transition energies: New approach to a comprehensive
  evaluation.
\newblock {\em Rev. Mod. Phys.} {\bf 2003}, {\em 75},~35--99.
\newblock {\url{https://doi.org/10.1103/RevModPhys.75.35}}.

\bibitem[Mendenhall et~al.(2019)Mendenhall, Hudson, Szabo, Henins, and
  Cline]{2019-Mo}
Mendenhall, M.H.; Hudson, L.T.; Szabo, C.I.; Henins, A.; Cline, J.P.
\newblock The molybdenum K-shell X-ray emission spectrum.
\newblock {\em J. Phys. At. Mol. Opt. Phys.}
  {\bf 2019}, {\em 52},~215004.
\newblock {\url{https://doi.org/10.1088/1361-6455/ab45d6}}.

\bibitem[Hudson et~al.(2020)Hudson, Cline, Henins, Mendenhall, and
  Szabo]{2020-SI}
Hudson, L.T.; Cline, J.P.; Henins, A.; Mendenhall, M.H.; Szabo, C.
\newblock Contemporary X-ray wavelength metrology and traceability.
\newblock {\em Radiat. Phys. Chem.} {\bf 2020}, {\em 167},~108392.
\newblock
  {\url{https://doi.org/10.1016/j.radphyschem.2019.108392}}.

\bibitem[Trassinelli et~al.(2016)Trassinelli, Anagnostopoulos, Borchert, Dax,
  Egger, Gotta, Hennebach, Indelicato, Liu, Manil, Nelms, Simons, and
  Wells]{2016-pion}
Trassinelli, M.; Anagnostopoulos, D. {F.}; Borchert, G.; Dax, A.; Egger,  {J.-P.;}
  Gotta, D.; Hennebach, M.; Indelicato, P.; Liu,  {Y.-W.;} Manil, B.;~et~al.
\newblock Measurement of the charged pion mass using X-ray spectroscopy of
  exotic atoms.
\newblock {\em Phys. Lett. B} {\bf 2016}, {\em 759},~583--588.
\newblock
  {\url{https://doi.org/10.1016/j.physletb.2016.06.025}}.

\bibitem[Okumura et~al.(2023)Okumura, Azuma, Bennett, Chiu, Doriese, Durkin,
  Fowler, Gard, Hashimoto, Hayakawa, Hilton, Ichinohe, Indelicato, Isobe,
  Kanda, Katsuragawa, Kawamura, Kino, Mine, Miyake, Morgan, Ninomiya, Noda,
  O'Neil, Okada, Okutsu, Paul, Reintsema, Schmidt, Shimomura, Strasser, Suda,
  Swetz, Takahashi, Takeda, Takeshita, Tampo, Tatsuno, Ueno, Ullom, Watanabe,
  and Yamada]{2023-Ne}
Okumura, T.; Azuma, T.; Bennett, D.A.; Chiu, I.; Doriese, W.B.; Durkin, M.S.;
  Fowler, J.W.; Gard, J.D.; Hashimoto, T.; Hayakawa, R.;~et~al.
\newblock Proof-of-principle experiment for testing strong-field quantum
  electrodynamics with exotic atoms: High precision x-ray spectroscopy of
  muonic neon.
\newblock {\em Phys. Rev. Lett.} {\bf 2023}, {\em 130},~173001.
\newblock {\url{https://doi.org/10.1103/PhysRevLett.130.173001}}.

\bibitem[Curceanu et~al.(2023)Curceanu, Abbene, Amsler, Bazzi, Bettelli,
  Borghi, Bosnar, Bragadireanu, Buttacavoli, Cargnelli, Carminati, Clozza,
  Deda, Del~Grande, De~Paolis, Dulski, Fiorini, Friščić, Guaraldo, Iliescu,
  Iwasaki, Khreptak, Manti, Marton, Miliucci, Moskal, Napolitano, Niedźwiecki,
  Onishi, Piscicchia, Principato, Sada, Scordo, Sgaramella, Shi, Silarski,
  Sirghi, Sirghi, Skurzok, Spallone, Toho, Tüchler, Doce, Yoshida, Zappettini,
  and Zmeskal]{2023-Daphne}
Curceanu, C.; Abbene, L.; Amsler, C.; Bazzi, M.; Bettelli, M.; Borghi, G.;
  Bosnar, D.; Bragadireanu, M.; Buttacavoli, A.; Cargnelli, M.;~et~al.
\newblock Kaonic atoms at the DA$\Phi$NE collider: A strangeness adventure.
\newblock {\em Front. Phys.} {\bf 2023}, {\em 11}, 1240250.
\newblock {\url{https://doi.org/10.3389/fphy.2023.1240250}}.
}}
\end{thebibliography}
\end{document}